\documentclass[aps,prl,twocolumn,groupedaddress,showpacs]{revtex4}

\usepackage{graphicx}
\usepackage{amsfonts}
\usepackage{amsmath}
\usepackage{amssymb}
\usepackage{bm}

\begin{document}
\title{A Topological Spin Chern Pump}

\author{C. Q. Zhou$^1$}
\author{Y. F. Zhang$^1$}
\author{L. Sheng$^1$}
\email{shengli@nju.edu.cn}
\author{R. Shen$^1$}
\author{D. N. Sheng$^2$}
\author{D. Y. Xing$^1$}
\email{dyxing@nju.edu.cn}
\affiliation{$^1$National Laboratory of Solid State Microstructures and
Department of Physics, Nanjing University, Nanjing 210093, China\\
$^2$ Department of Physics and Astronomy, California State
University, Northridge, California 91330, USA}
\date{\today }

\begin{abstract}
We propose a one-dimensional electron model with parameters
modulated adiabatically in closed cycles, which can
continuously pump spin to leads. By defining the spin-polarized
Wannier functions, we show that the spin pump is protected by the spin Chern numbers,
so that it is stable to perturbations violating the time-reversal symmetry
and spin conservation.  Our work demonstrates
the possibility and principle to realize topological
spin pumps independent of any symmetries, and also
suggests a possible way to experimentally
observe the bulk topological invariants.

\end{abstract}

 \maketitle

The quantum Hall (QH) effect discovered in 1980~\cite{Klitzing} is
the first example of topological state
in the field of condensed matter physics. 
Since then, there has been continuously strong interest in
topological phenomena of condensed matter systems.
Laughlin~\cite{Laughlin} interpreted the integer QH effec 
as a quantum charge pump. 
Increasing the magnetic flux by a single flux quantum that threads a looped QH
ribbon constitutes a cycle of the pump
due to gauge invariance, transferring an integer-quantized amount of charge
from one edge of the ribbon to the other.
Thouless, Kohmoto, Nightingale, and Nijs~\cite{Thouless0}
showed that the QH state can be classified by a topological
invariant, the Chern number.
Thouless and Niu~\cite{Thouless1,Thouless2} also
established a general relation between the Chern number and the charge
pumped during a period of slow variation of potential in the
Schr\"{o}dinger equation.

Recently, an important discovery was the
topological insulator,~\cite{TI1,TI2,TI3,TI4}
a new quantum state of matter existing
in nature. Different from the QH systems, the topological
insulators preserve the time-reversal (TR) symmetry.
Two-dimensional topological insulators,
also called the quantum spin Hall (QSH) systems,
have a bulk band gap and a pair of gapless
helical edge states traversing the bulk gap.
When electron spin is conserved, the topological properties
of the QSH systems can be easily understood, as a QSH system
can be viewed as two independent QH systems without Landau levels.~\cite{Haldane}
When the spin conservation is destroyed,
unconventional topological invariants are needed to
classify the QSH systems. The $Z_2$ index~\cite{Z2index} and
the spin Chern numbers~\cite{spinch1,spinch2,spinch3}
have been proposed to describe the QSH systems.
While the two different invariants are found to be equivalent to
each other for TR-invariant systems,~\cite{spinch2,spinch3} they lead to
controversial predictions when the TR symmetry is broken.
The definition of the $Z_2$ index explicitly
relies on the presence of TR symmetry,
suggesting that the QSH state turns into a trivial insulator
once the TR symmetry is broken.
However, calculations~\cite{spinch4} based upon
the spin Chern numbers showed that the nontrivial topological
properties of the QSH systems remain
intact when the TR symmetry is broken, as long as
the band gap and spin spectrum gap stay open.
The nonzero spin Chern numbers
guarantee that the edge states must appear on the sample boundary,~\cite{spinch5} which
could be either gaped or gapless, depending on symmetries or  spatial
distributions of the edge states.~\cite{spinch6}
This prediction was supported
by the recent experimental observation of the QSH effect in InAs/GaSb bilayers
under broken TR symmetry.~\cite{RRDu}

Spin pumps promise broad applications in spintronics, e.g.,
the resulting spin battery is the spintronic analog of the charge
battery in conventional electronics.
Topological spin pumps~\cite{Sharma,Shindou,Z2pump}
are expected to have an advantage over other approaches,
~\cite{sPump1,sPump2,sPump3,sPump4,sPump5,sPump6}
being insusceptible to environmental perturbations.
When spin $s_z$ is conserved, the idea of the Thouless charge pump
was extended to construct quantized adiabatic spin pumps.~\cite{Sharma,Shindou}
However, Fu and Kane argued~\cite{Z2pump} that, unlike the charge, the
spin does not obey a fundamental conservation law,
and they introduced a more general concept of the $Z_2$ pump.
In the $Z_{2}$ pump, while the amount of spin pumped
per cycle is not integer-quantized
in the absence of spin conservation, the pumping process is protected
by a $Z_2$ topological invariant, provided that
the TR symmetry is present. So far, in existing
proposals, either spin conservation
or TR symmetry is necessary for constructing
topological spin pumps, which greatly restricts their practical applications.
Robust spin pumps protected by topology alone, independent of any symmetries,
are still awaited.

In this Letter, we predict another intriguing effect resulting from the
spin Chern numbers, namely, topological spin pumping. 
A one-dimensional electron model with parameters
modulated adiabatically in closed cycles is proposed, which can
continuously pump spin into leads.
By defining the spin-polarized Wannier functions (SPWFs), we reveal that
the spin pumping effect is a direct manifestation of the
nontrivial topological properties of the electron wavefunctions,
characterized by nonzero spin Chern numbers.
In contrast to the $Z_2$ pump,
this spin Chern pump remains to be robust
in the presence of magnetic impurities,
which destroy both the TR symmetry
and spin conservation.
Our work demonstrates the possibility and principle to
implement robust topological spin pumps independent of any
symmetries, and also
suggests a possible way to observe the bulk topological invariants experimentally.

\begin{figure}
\includegraphics[width=2.2in]{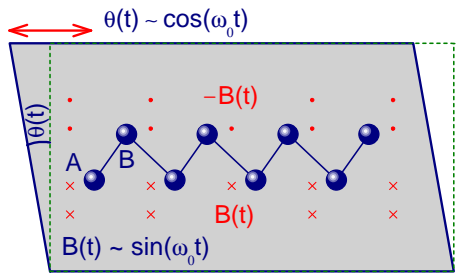}
\caption{
A zigzag chain of atoms are affixed on a substrate.
Oscillatory shear deformation of the substrate is stimulated,
resulting in a periodic modulation of hopping
integrals between nearest-neighboring atomic sites. A nonuniform oscillating magnetic field
is applied in such a manner that electrons on the $A$ and $B$ sites
experience opposite Zeeman fields at any time.} \label{Sch}
  \end{figure}
	
We consider a one-dimensional electron model with Hamiltonian~\cite{note0}
\begin{equation}
H_{P}=\sum_{\langle i,j\rangle}t_{i,j}c_{i}^\dagger c_{j}+g(t)\sum_{i}(-1)^{i}
c_{i}^\dagger s_{z}c_{i}\ ,
\label{H_TBA}
\end{equation}
where $c_{i}^{\dagger}=(c_{i\uparrow}^\dagger,c_{i\downarrow}^{\dagger})$
are the creation operators in the spinor representation for electrons
with up and down spins on site $i$, $t_{i,j}$ is the periodically
varying hopping integral between the nearest neighboring sites,
given by $t_{i,i+1}=t_{i+1,i}=t_{0}+(-)t_{1}\cos\omega_0t$
for $i$ on the $A$ ($B$) sublattices, $g(t)=g_{0}\sin\omega_{0}t$
is the Zeeman splitting energy, and $s_z$ is the
Pauli matrix acting on the electron spin.
A possible experimental realization of this model is illustrated in Fig.\ 1.
It is easy to see that Eq.\ (1) preserves the TR symmetry, i.e.,
$H_{P}(-t)=\Theta H_{P}(t)\Theta$ with $\Theta$ as the ordinary TR operator.
For an infinitely long chain of atoms, the eigenenergies of Eq.\ (\ref{H_TBA})
can be obtained by the Fourier transform, yielding
\begin{equation}
E(k_x)=\pm\sqrt{g^2(t)+4t^2_{0}\cos^2\frac{k_xa_{0}}{2\hbar}+
\alpha^2(t)\sin^2\frac{k_xa_{0}}{2\hbar}}
\label{Ekx}
\end{equation}
with $\alpha(t)=2t_{1}\cos\omega_{0}t$ and $a_{0}$ the lattice constant. 
Given $t_{0}\gg t_{1},g_{0}>0$, the system has
a middle band gap between $\pm\sqrt{\alpha^2(t)+g^2(t)}$, which is finite at any time. 
In the adiabatic limit, on the torus of $k_x$ and $t$, one can
define the spin Chern numbers $C_{\pm}$ in a standard way,~\cite{spinch2,spinch3,spinch4}
and obtain $C_{\pm}=\pm 1$.

In what follows we want to set up a relation of the 
nontrivial topological properties
of the system to the spectral flow of the centers of 
mass of the SPWFs. In order to show the robustness
of nontrivial topological properties, we introduce magnetic impurities
with randomly oriented classical spins into the system. The Hamiltonian is given by
$H_{I}=V_{0}\sum_{\alpha}c_{\alpha}^{\dagger}
{\bf s}\cdot {\bf m}_{\alpha}c_{\alpha}$,
where $\alpha$ runs over all the impurity sites, and ${\bf m}_{\alpha}$
is a unit vector in the direction of the $\alpha$-th impurity spin.
Apparently, the presence of the magnetic disorder destroys both the spin conservation
and TR symmetry of the system. 

\begin{figure}
\includegraphics[width=2.2in]{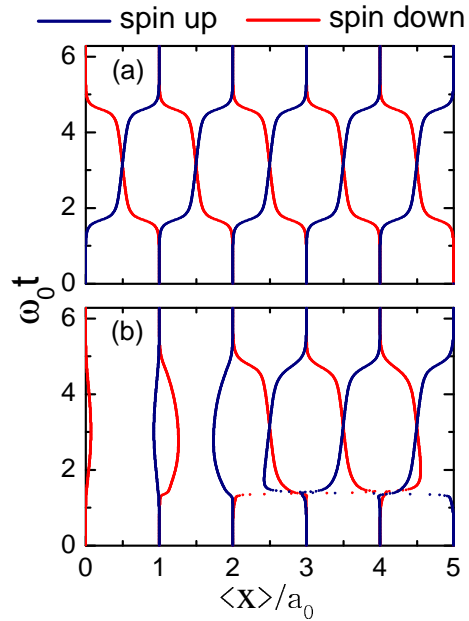}
\caption{Evolution of centers of the Wannier functions for
the spin-up and spin-down sectors. For clarity, only 5 unit cells of a long
chain with periodic boundary condition are displayed, in which
$20\%$ of the atoms are assumed to be
replaced with magnetic impurities. The other parameters are taken as
$t_{1}=g_{0}=0.1$, and $V_{0}=1$ (a) and  $V_{0}=2.5$ (b), with $t_{0}$
as the unit of energy.
} \label{Fig1}
  \end{figure}
We diagonalize the total Hamiltonian $H_{P}+H_{I}$ numerically,
and the eigenenergies and eigenstates are denoted as $E_n$ and
$\vert\varphi_{n}\rangle$. By using the same procedure as calculating the spin Chern
numbers, the occupied valence bands can be partitioned into two spin
sectors by diagonalizing the projected spin operator $Ps_zP$ with $P=
\sum_{E_{n}<E_{\mbox{\tiny F}}}\vert\varphi_{n}\rangle\langle
\varphi_{n}\vert$ as the projection operator
to the occupied space. If the spin is conserved, the eigenvalues
of $Ps_zP$ have only two values:  $1$ or $-1$. When the spin conservation
is broken weakly, there is still a finite gap in the eigen-spectrum of $Ps_{z}P$,
which naturally divides the spectrum into two sectors:  spin-up and spin-down sectors.
The eigenstates of $Ps_zP$ for the two spin sectors are denoted by $\vert{\psi}_{m\pm}\rangle$.
By definition, $\vert{\psi}_{m\pm}\rangle$ are essentially the maximally spin-polarized
states. Then we can construct the Wannier functions~\cite{Wannier,QiXL} $\vert \chi_{m\pm}\rangle$
for the spin-up and spin-down sectors,
respectively, which are called the SPWFs.

The evolution of the centers of mass $\langle x\rangle$
of the SPWFs with time is plotted in
Fig.\ 2 for two different disorder
strengths. It is found that for relatively weak magnetic disorder,
all the centers of $\vert\chi_{m+}\rangle$
move rightwards, 
each center on average shifting a lattice constant per cycle,
and those of $\vert\chi_{m-}\rangle$ move in the opposite direction,
as shown in Fig.\ 2(a).
According to the general theory,~\cite{QiXL}
the total displacement of the centers of
$\vert\chi_{m+}\rangle$ ($\vert\chi_{m-}\rangle$)
per cycle divided by the length of the system  is equal to the
spin Chern number $C_{+}=1$ ($C_{-}=-1$)
for the spin-up (spin-down) sector.
For strong magnetic disorder, the ordered movement of the Wannier
centers is interrupted, though rearrangement of some centers still happens locally,
as shown in Fig.\ 2(b), indicating that the system becomes
topologically trivial ($C_{\pm}=0$).

It is worth pointing out that the space spanned
by the SPWFs $\vert\chi_{m\pm}\rangle$ is identical to that spanned by
$\vert\varphi_{n}\rangle$ for $E_n<E_{\mbox{\mbox{\tiny F}}}$, namely,
$\sum_{m}\vert\chi_{m+}\rangle\langle\chi_{m+}\vert+
\sum_{m}\vert\chi_{m-}\rangle\langle\chi_{m-}\vert= P$. The SPWFs are 
just another equivalent representation of the occupied space.
Therefore, the $N$ electrons occupying the energy eigenstates
$\vert\varphi_{n}\rangle$ for $E_n<E_{\mbox{\mbox{\tiny F}}}$ may 
also be equivalently considered as two groups:
$N/2$ electrons occupying $\vert\chi_{m+}\rangle$ and
$N/2$ electrons occupying $\vert\chi_{m-}\rangle$.
The counter flows of the centers of the SPWFs
observed in Fig.\ 2(a), as a consequence of the nonzero spin Chern numbers,
represent the true movements of the electrons in
the spin-up and spin-down sectors with time.
Without the TR symmetry and spin conservation,
such nontrivial spectral flows become visible only if the occupied space
is properly partitioned, as has been done above.
It is expected that if leads are strongly connected 
to the two ends of the atomic chain,
the opposite movements of the electrons in the spin-up and spin-down sectors 
will extend into the leads, transferring spin to the leads continuously.
The system becomes a topological spin pump.

We now consider a sufficiently long pump for
$x<0$ and a lead for $x>0$, which are in good contact with each other.
In order to obtain a transparent analytical expression for the pumped spin,
we expand the Hamiltonian Eq.\ (\ref{H_TBA}) 
for the pump around $k=\pi\hbar/a$,
where the band gap is minimal, yielding
\begin{equation}
H_{P}=\alpha(t)\sigma_x+v_{\mbox{\tiny F}}
p_{x}\sigma_y+g(t)s_z\sigma_z\ ,
\end{equation}
where $v_{\mbox{\tiny F}}=t_{0}a_{0}/\hbar$, $p_x=k_x-\pi\hbar/a_{0}$, and $\sigma_{x(y,z)}$
are the Pauli matrices associated with the $AB$ sublattices.
The Hamiltonian for the lead is taken to be
\begin{equation}
H_{L}=v_{\mbox{\tiny F}}p_{x}\sigma_y\ .
\end{equation}
The Fermi level is set
to be $E_{\mbox{\mbox{\tiny F}}}=0$, inside
the bulk band gap of the pump. It is assumed that
within the decay length of electron wavefunctions into the pump,
there exists only one magnetic impurity at $x=0$  with
the potential taken as $H_{I}=V(x)s_{x}$. Here,
$V(x)$ is modeled as a square potential centered at $x=0$ with height $V_{0}$
and width $d$. If $d$ is much smaller than
the decay length of wavefunctions,
by taking the $d\rightarrow 0$ limit and keeping $U_{0}=V_{0}d$
finite, it can be shown that the scattering effect of the impurity
potential is equivalent to imposing a unitary boundary condition
for the electron wavefunctions
\begin{equation}
\Psi(x=0^{+})=S\Psi(x=0^{-})\ , 
\end{equation}
where $S=e^{-i\phi\sigma_ys_x}$ with
$\phi=U_{0}/\hbar v_{\mbox{\tiny F}}$. Without the impurity ($\phi=0$),
Eq.\ (5) will reduce to the ordinary continuity
condition $\Psi(x=0^{+})=\Psi(x=0^{-})$.

Calculation of the spin pumped into the lead per cycle amounts to
solving the scattering problem of an electron incident at the Fermi level
from the lead.~\cite{SMat1,SMat2} 
We first consider the case, where the spin of the incident electron is parallel
to the $z$ axis. On the bases $(|\uparrow,1\rangle,$ $
\vert\uparrow,-1\rangle,$ $\vert\downarrow,1\rangle$, $\vert\downarrow,-1\rangle)$
with the kets as the eigenstates of $s_z$ and
$\sigma_z$, the wavefunction in the lead is given by
\begin{equation}
\Psi(x)=\frac{1}{\sqrt{2}}\left(\begin{array}{c}
1\\
-i\\
0\\
0
\end{array}
\right)
+\frac{r_{\uparrow\uparrow}}{\sqrt{2}}
\left(\begin{array}{c}
1\\
i\\
0\\
0
\end{array}
\right)
+\frac{r_{\downarrow\uparrow}}{\sqrt{2}}
\left(\begin{array}{c}
0\\
0\\
1\\
i
\end{array}
\right)\ ,
\end{equation}
for $x>0$, and that in the pump is given by
\begin{equation}
\Psi(x)=C_{1}
\left(\begin{array}{c}
\sin\frac{\varphi}{2}\\
\cos\frac{\varphi}{2}\\
0\\
0
\end{array}
\right)e^{\gamma x}
+C_{2}
\left(\begin{array}{c}
0\\
0\\
-\sin\frac{\varphi}{2}\\
\cos\frac{\varphi}{2}
\end{array}\right)e^{\gamma x}\ ,
\end{equation}
for $x<0$. Here, $\varphi=\mbox{Arg}[\alpha(t)+ig(t)]$,
and $\gamma=\sqrt{\alpha^2(t)+g^2(t)}/\hbar v_{F}$.
By substituting Eqs.\ (6) and (7) into Eq.\ (5),
it is straightforward to derive for the reflection amplitudes
$r_{\uparrow\uparrow}=-[\cos(2\phi)\cos(\varphi)+i\sin(\varphi)]$ and
$r_{\downarrow\uparrow}=i\sin(2\phi)\cos(\varphi)$.
Similarly, by considering the case, where the spin of the incident
electron is antiparallel to
the $z$ axis, one can obtain
$r_{\downarrow\downarrow}=r_{\uparrow\uparrow}^{*}$ and
$r_{\uparrow\downarrow}=r_{\downarrow\uparrow}$.

The $z$-component of the pumped spin 
per cycle in unit of $\hbar/2$ is given by~\cite{SMat1,SMat2}
\begin{eqnarray}
\Delta s_{z}&=&\frac{1}{2\pi i}\oint_{T} 
dt\Bigl(r^{*}_{\uparrow\uparrow}\frac{dr_{\uparrow\uparrow}}{dt}
-r^{*}_{\downarrow\downarrow}\frac{dr_{\downarrow\downarrow}}{dt}\nonumber\\
&-&r^{*}_{\downarrow\uparrow}\frac{dr_{\downarrow\uparrow}}{dt}+
r^{*}_{\uparrow\downarrow}\frac{dr_{\uparrow\downarrow}}{dt}\Bigr)\ ,
\end{eqnarray}
with $T=2\pi/\omega_{0}$ as a period of the pump.
Here, the third and fourth terms in the integrand have no contribution, since
$r_{\uparrow\downarrow}=r_{\downarrow\uparrow}$ is always imaginary.
Due to $r_{\downarrow\downarrow}=r_{\uparrow\uparrow}^{*}$, the first and
second terms make an equal contribution. Therefore,
$\Delta s_{z}=\frac{1}{\pi i}\oint_{T} r^{*}_{\uparrow\uparrow}dr_{\uparrow\uparrow}$,
which can be further evaluated to be
\begin{equation}
 \Delta s_z =  2-4\phi^2 + {\cal O}(\phi^4)\ ,
\end{equation}
for $\phi \ll 1$.
Similarly, one can find $\Delta s_{x}=\Delta s_y=0$. In Eq.\ (9),
 $\Delta s_{z}$ is quantized to be $2$ at $\phi=0$, and there is
a small deviation from the quantized value for small $\phi$, being
consistent with the analysis of the SPWFs. The small deviation arises from
the destruction of the spin conservation by the magnetic impurity,
rather than the breaking of the TR symmetry.
Such a deviation occurs as well in the absence of the magnetic impurity, if the
Rashba spin-orbit coupling is included,~\cite{Z2pump}
which destroys the spin conservation but preserves the TR symmetry.
Physically, it is because the electron wavefunction $\Psi(x)$, given by Eqs.\ (6) and (7),
is not an eigenstate of $s_{z}$ when $\phi\neq 0$. Moreover, the direction
of the spin polarization of the wavefunction 
varies with time ($\varphi$ is a function
of $t$), and so the
quantized value cannot be recovered
by properly choosing the spin quantization axis.

\begin{figure}
\includegraphics[width=2.3in]{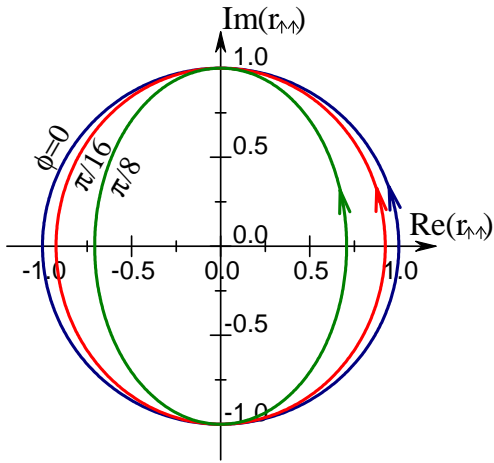}
\caption{Trajectories of $r_{\uparrow\uparrow}$ on the complex plane for three
different strengths of impurity scattering potential.} \label{Fig2}
  \end{figure}

To get some more insight into the spin pump,
we plot the trajectories of $r_{\uparrow\uparrow}$ in a cycle
on the complex plane in Fig.\ 3, for three different $\phi$.
Each trajectory is a closed orbit simply because the Hamiltonian
is periodic in time. At $\phi=0$,
the orbit of $r_{\uparrow\uparrow}$ is a unit circle, and increasing
$\phi$ deforms the orbit. In general,
one can find that $\Delta s_z$ equals to the area enclosed by
the trajectory of $r_{\uparrow\uparrow}$ divided by $\pi/2$.
Therefore, any small perturbation may cause a small deformation
of the trajectory from the unit circle, but can neither stop the spin pumping 
nor change the sign of $\Delta s_{z}$.
This reflects the topological stability of the spin pump
from another aspect.

Finally, we wish to make a comment on the $Z_{2}$ spin pump proposed by Fu and Kane.~\cite{Z2pump}
These authors studied the same model as Eq.\ (\ref{H_TBA}), and showed that
for a system with closed ends, there are a pair of bound states localized near
each end with energy levels crossing the bulk energy gap with time going on. 
The end states
exhibit level crossing
at the TR invariant point $t=T/2$, forming a Kramers doublet.
For the system ``weakly'' coupled to leads, there occurs a resonance in 
reflection amplitudes, when the Kramers degenerate end states appear. Such a
resonance structure allows spin to be pumped into the leads.
However, if the TR symmetry is broken, an energy gap will open
at the level-crossing point, and the spin pumping will be stopped.
Therefore, they concluded that
the TR symmetry plays a crucial role in the $Z_2$ spin pump.
Apparently, the $Z_2$ pump essentially reveals the properties of the
end states.  An important difference of the present spin Chern pump
 from the $Z_2$
pump is the ``strong'' connection between the pump and leads, which allows
the electrons to move freely between the pump and leads,
without appearance and participation of the end states.
The spin pumping process in the spin Chern pump
is guaranteed by the spin Chern numbers alone,
and hence robust against symmetry-breaking
perturbations, as has been shown above. 
It reveals the bulk topological invariants directly.

This work was supported by the State Key Program for Basic Researches of China under
grants numbers 2014CB921103 (LS), 2011CB922103 and 2010CB923400 (DYX), the National
Natural Science Foundation of China under grant numbers 11225420 (LS),
11174125, 91021003 (DYX) and a project funded by the PAPD of Jiangsu Higher
Education Institutions. We also thank the US NSF grants numbers DMR-0906816 and DMR-
1205734 (DNS).


\begin{thebibliography}{99}
\bibitem{Klitzing} K. Klitzing, G. Dorda, M, Pepper,  Phys. Rev. Lett. {\bf 45}, 494 (1980).
\bibitem{Laughlin} R. B. Laughlin, Phys. Rev. B {\bf 23}, 5632 (1981).
\bibitem{Thouless0} D. J. Thouless, M. Kohmoto, M. P. Nightingale, and M. den Nijs,
Phys. Rev. Lett. 49, 405 (1982).
\bibitem{Thouless1} D. J. Thouless, Phys. Rev. B {\bf 27}, 6083 (1983).
\bibitem{Thouless2} Q. Niu and D. J. Thouless, J. Phys. A {\bf 17}, 2453 (1984).
\bibitem{TI1} C. L. Kane and E. J. Mele, Phys. Rev. Lett. {\bf 95}, 226801
(2005);
 B. A. Bernevig, and S. C. Zhang, Phys. Rev. Lett. {\bf 96},
106802 (2006).
\bibitem{TI2} J. E. Moore, and L. Balents, Phys. Rev. B {\bf 75}, 121306
(R) (2007);
L. Fu and C. L. Kane, Phys. Rev. B {\bf 76}, 045302 (2007); L.
Fu, C. L. Kane, E. J. Mele, Phys. Rev. Lett. {\bf 98}, 106803
(2007).
\bibitem{TI3} M. Z. Hasan and C. L. Kane, Rev. Mod. Phys. {\bf 82}, 3045
(2010).
\bibitem{TI4} X. L. Qi and S. C. Zhang, Physics Today {\bf 63}, 33 (2010).
\bibitem{Haldane} F. D. M. Haldane, Phys. Rev. Lett. {\bf 61}, 2015 (1988).
\bibitem{Z2index} C. L. Kane, and E. J. Mele, Phys. Rev. Lett. {\bf 95}, 146802 (2005).
\bibitem{spinch1} D. N. Sheng, Z. Y. Weng, L. Sheng, and F. D. M. Haldane,
Phys. Rev. Lett. {\bf 97}, 036808 (2006).
\bibitem{spinch2} E. Prodan, Phys. Rev. B \textbf{80}, 125327 (2009); E. Prodan, New J. Phys. \textbf{12}, 065003 (2010).
\bibitem{spinch3} H. C. Li, L. Sheng, D. N. Sheng, and D. Y. Xing, Phys. Rev. B \textbf{82},
165104 (2010).
\bibitem{spinch4} Y. Yang, Z. Xu, L. Sheng, B. G. Wang, D. Y. Xing, and D. N. Sheng
Phys. Rev. Lett. {\bf 107}, 066602 (2011).
\bibitem{spinch5} H. C. Li, L. Sheng, and
D.Y. Xing, Phys. Rev. Lett. {\bf 108}, 196806 (2012).
\bibitem{spinch6} H. C. Li, L. Sheng, R. Shen, L. B. Shao, B. G. Wang, D. N. Sheng, and
D. Y. Xing, Phys. Rev. Lett. {\bf 110}, 266802 (2013).
\bibitem{RRDu} L. Du, I. Knez, G. Sullivan, R.-R. Du, cond-mat/13061925 (2013).
\bibitem{Sharma} P. Sharma and C. Chamon, Phys. Rev. Lett. {\bf 87}, 096401 (2001).
\bibitem{Shindou} R. Shindou, J. Phys. Soc. Jpn. {\bf 74}, 1214 (2005).
\bibitem{Z2pump} L. Fu and C. L. Kane, Phys. Rev. B {\bf 74}, 195312 (2006).

\bibitem{sPump1} E. R. Mucciolo, C. Chamon, and C. M. Marcus, Phys. Rev. Lett.
{\bf 89}, 146802 (2002).
\bibitem{sPump2} T. Aono, Phys. Rev. B {\bf 67}, 155303 (2003).
\bibitem{sPump3} M. Governale and F. T. R. Fazio, Phys. Rev. B
{\bf 68}, 155324 (2003).
\bibitem{sPump4} R. Citro and F. Romeo, Phys. Rev. B
{\bf 73}, 233304 (2006).
\bibitem{sPump5} A. Schiller and A. Silva, Phys. Rev. B {\bf 77}, 045330 (2008).
\bibitem{sPump6} C. Sandweg $et$ $al$, Phys. Rev. Lett. {\bf 106}, 216601 (2011).
\bibitem{note0} Fu and Kane studied the same model, focusing on the
limit of weak coupling to leads.~\cite{Z2pump} We will consider the case of strong coupling,
where the physics is found to be quite different.
\bibitem{Wannier} G. H. Wannier, Rev. Mod. Phys. {\bf 34}, 645 (1962).
\bibitem{QiXL} X. L. Qi, Phys. Rev. Lett. {\bf 107}, 126803 (2011).
\bibitem{SMat1} M. B\"{u}ttiker, H. Thomas, A. Pr\^{e}tre,
Z. Phys. B {\bf 94}, 133 (1994).
\bibitem{SMat2} P.W. Brouwer, Phys. Rev. B {\bf 58}, 10135
(1998).



\end{thebibliography}
\end{document}